\documentclass[twocolumn]{aastex631}

\renewcommand\thesubsection{\thesection.\alph{subsection}}

\usepackage{natbib}

\newcommand{\elsasser}{Els\"asser\ }
\received{datum, 2021}
\revised{datum, 2021}
\accepted{datum, 2021}

\shorttitle{Weak damping of propagating kink waves}
\shortauthors{Morton et al.}

\begin{document}

\title{Weak damping of propagating MHD kink waves in the quiescent corona}

\correspondingauthor{Richard J. Morton}
\email{richard.morton@northumbria.ac.uk}
\author[0000-0001-5678-9002]{Richard J. Morton}
\affil{Northumbria University, Newcastle upon Tyne, NE1 8ST, UK}

\author[0000-0001-6021-8712]{Ajay K. Tiwari}
\affiliation{Northumbria University, Newcastle upon Tyne, NE1 8ST, UK}
\affiliation{Centrum Wiskunde \& Informatica, Amsterdam, The Netherlands}

\author[0000-0001-9628-4113]{Tom Van Doorsselaere}
\affiliation{Centre for mathematical Plasma Astrophysics, Department of Mathematics, KU Leuven, Celestijnenlaan 200B bus 2400, B-3001 Leuven, Belgium}

\author[0000-0002-7863-624X]{James A. McLaughlin}
\affiliation{Northumbria University, Newcastle upon Tyne, NE1 8ST, UK}

\begin{abstract}
Propagating transverse waves are thought to be a key transporter of Poynting flux throughout the Sun's atmosphere. Recent studies have shown that these transverse motions, interpreted as the magnetohydrodynamic kink mode, are prevalent throughout the corona. The associated energy estimates suggest the waves carry enough energy to meet the demands of the coronal radiative losses in the quiescent Sun. However, it is still unclear how the waves deposit their energy into the coronal plasma. We present the results from a large-scale study of propagating kink waves in the quiescent corona using data from the Coronal Multi-channel Polarimeter (CoMP). The analysis reveals that the kink waves appear to be weakly damped, which would imply low rates of energy transfer from the large-scale transverse motions to smaller-scales via either uni-turbulence or resonant absorption. This raises questions about how the observed kink modes would deposit their energy into the coronal plasma. Moreover, these observations, combined with the results of Monte Carlo simulations, lead us to infer that the solar corona displays a spectrum of density ratios, with a smaller density ratio (relative to the ambient corona) in quiescent coronal loops and a higher density ratio in active region coronal loops.

\end{abstract}

\keywords{Sun: corona  ---  waves  --- magnetohydrodynamics (MHD) }

\section{Introduction}\label{intro}

It is now well established that transverse wave modes are ubiquitous throughout the Sun's atmosphere. The most unambiguous signature of these waves is the 
transverse displacement of magnetised wave guides, observed in both the chromosphere (e.g., fibrils \citealt{MORetal2012, MORetal2014, JAFetal2016}; super-penumbral fibrils \citealt{MORetal2021}; spicules \citealt{OKADEP2011}) and corona 
\citep[e.g.,][]{TOMetal2007, MCIetal2011, THUetal2014, MORetal2019, YANGetal2020}. This motion has been interpreted from magnetohydrodynamic (MHD) wave theory as the kink mode \citep[e.g.,][]{VANetal2008b}. Within the corona, there appear to be three common variants of the kink mode identified: the rapidly-damped standing mode 
\citep[e.g.,][]{ASCetal1999,  NAKetal1999, GODetal2016}; the decayless standing mode \citep[e.g.,][]{NISetal2013, ANFetal2015, KARVAND2020}
and
the propagating mode \citep{TOMMCI2009, MORetal2015, MORetal2016}. The 
propagating kink mode is found throughout the corona \citep{MORetal2019,YANGetal2020} and energy estimates suggest the waves are powerful enough to meet the heating rate required to balance the radiative losses in the 
quiescent corona and coronal holes \citep{MCIetal2011, WEBetal2018}. On the other hand, the damped standing mode is excited by impulsive 
events and occurs too infrequently to provide a 
significant energetic contribution \citep{TERARR2018,Nakariakov_2020}.  For the decay-less variant, the situation is less clear: they have only been detected in active regions, but \citet{Hillier_2020} find that their energy content is potentially insufficient to meet active region heating requirements. The 
connection between the decay-less standing mode and the propagating kink mode, if any, is still unclear. Interestingly, the nature of some of the events classified as standing kink modes has been called into question \citep{Hindman_2014,Jain_2015}, with the suggestion that a number of the observed cases may arise as a response of a magnetic arcade to a fast wave disturbance propagating obliquely to the magnetic field.

\section{Kink waves}

In order to provide the necessary context for our later analysis and discussions, we provide a summary of some of the key observational and theoretical results
for kink modes. While the new results presented later are focused the propagating kink modes, we draw comparisons to the standing modes.

\subsection{Propagating kink waves}
The propagating kink waves have largely been observed with Doppler velocity measurements of the 1074.7~nm Fe XIII line 
from the Coronal Multi-Channel Polarimeter \citep[CoMP -][]{TOMetal2008}, with recent studies clearly showing they 
exist throughout the corona \citep{MORetal2019,YANGetal2020}. It appears that the waves propagate along coronal structures (e.g., loops, plumes) that are over-dense compared to the ambient plasma, with the over-density signified by enhanced emission in coronal EUV emission lines. The waves propagate at speeds between $300-700$~ km/s \citep{MORetal2016,YANGetal2020}, which is significantly greater than the typical coronal sound speed. The power spectra of velocity fluctuations demonstrate power law behaviour over the currently observable frequency range \citep[$10^{-4}-10^{-2}$~Hz,][]{MORetal2016}. There is also an enhancement of the wave power centered on 4~mHz that has a direct correspondence to the transverse displacements of coronal
structures observed with the Solar Dynamics Observatory (SDO) \citep{MORetal2019}.  

The modelling of MHD waves in coronal structures is typically simplified to assume an over-dense wave-guide with cylindrical cross-sectional geometry \citep[e.g.,][]{SPR1982,EDWROB1983}. The velocity fluctuations observed in CoMP (and transverse displacements observed in SDO) are then interpreted as the kink mode. In the long wavelength limit, the kink mode is Alfv\'enic in nature \citep{GOOetal2009,GOOetal2012} and the phase speed is given by
\begin{equation}
c_k^2 = \frac{B_i^2+B_e^2}{\mu_0(\rho_i+\rho_e)},    
\end{equation}
which depends upon the local magnetic field, $B$, and density, $\rho$. The subscripts refer to the internal, $i$, and external, $e$, plasma quantities, and $\mu_0$ is the magnetic permeability.
The long wavelength regime occurs when $kR\ll 1$, where $k=2\pi/\lambda$ is the wavenumber ($\lambda$ the wavelength) of the wave and $R$ is the radius of the wave-guide\footnote{For clarity, we assume that density of the wave guide is inhomogeneous but smoothly varying perpendicular to the magnetic field. Then $R$ is such defined as being at the location where $\rho(R)$ is equal to density averaged over $R$.}. Given that the typical radius of coronal loops is $\sim 1$~Mm \citep{BROetal2013,Williams_2020} and the wavelengths for the propagating kink modes are $\lambda = c_k/f= 30-7000$~Mm (where $f$ is the frequency), the observed kink waves are comfortably in the long wavelength regime.

\subsection{Wave damping theory}
If the {{propagating}} kink modes are to play a relevant role in heating the corona, the waves have to dissipate the energy 
they transport into the plasma. In order for this to occur, the energy is required to be transferred to smaller spatial 
scales than those currently associated with the observed waves. A mechanism that has received considerable renewed interest recently is phase mixing \citep{Pritchett_1978,HEYPRI1983}.

The density contrast between the internal and external plasma of the coronal structures leads to a gradient in the Alfv\'en speed across the wave-guide boundary. The spatially-varying Alfv\'en frequency in the boundary layer enables phase mixing of rotational motions, which are described by Alfv\'en waves, and hence creates small spatial scales that eventually enables dissipation of the wave energy \citep[e.g.,][]{SOLetal2015,Pagano_2017,Pagano_2020}. Moreover, a resonance is created at the locations where the Alfv\'en frequencies in the boundary layer match that of the global kink frequencies. A consequence of the resonance is the energy in the transverse motions is transferred to rotational motions via resonant absorption \citep[e.g.,][]{ION1978,RUDROB2002, TERetal2010c,Goossens_2011}, and effectively couples the kink and Alfv\'en modes \citep{Pascoe_2010,PASetal2012}. 

The rate of damping for the propagating kink mode via resonant absorption in the thin boundary limit is given by \citep[see, e.g.,][]{TERetal2010c}\footnote{Similarly, for a standing kink mode the left-hand-side is replaced by the temporal rate of damping $\tau_\mathrm{RA}/P$, where $\tau_\mathrm{RA}$ is the damping rate and $P$ is the period \citep{RUDROB2002}.},
\begin{equation}
    \frac{L_D}{\lambda}=\frac{2}{\pi}\frac{R}{l}\frac{\rho_i+\rho_e}{\rho_i-\rho_e}.
    \label{eq:equib}
\end{equation}
Here, $L_D$ is the spatial damping length, and $l$ is the length-scale of inhomogeneous boundary layer. The factor $2/\pi$ comes from assuming a sinusoidal density profile across the boundary layer.  
The resonant damping of kink modes can be viewed as a linear mechanism for energy transfer, in that sense that it occurs in linearised MHD. This is readily seen in 
Eq.~\ref{eq:equib} where the damping is independent of amplitude.
\medskip

More recently it has been suggested that large amplitude kink waves can be damped by a non-linear mechanism, the 
so-called
`uni-turbulence' \citep{Magyar_2017,Magyar_2019,Van_Doorsselaere_2020} that causes a self-cascade of the wave energy. 
\citet{Van_Doorsselaere_2020} has computed the 
non-linear evolution of standing and propagating kink waves with \elsasser variables, in order to understand their non-linear damping as a consequence of the Kelvin-Helmholtz instability (KHI - see, e.g.,  \citealt{TERetal2008b,antolin2014,magyar2016}) or uni-turbulence \citep{Magyar_2017,Magyar_2019}. They predicted that damping times can be the same order of magnitude as those expected from resonant absorption for propagating waves. As with resonant absorption (and subsequent phase mixing), the key form of the inhomogeneity for uni-turbulence is related to the presence of field-aligned density enhancements (i.e., over-dense wave guides) and is suggested to be a form of generalised phase-mixing \citep{Magyar_2019}.

The time-scale for the energy cascade rate, $\tau_\mathrm{Pr}$, in propagating waves is given by \citep{Van_Doorsselaere_2020}, and recasting their damping for energy to velocity damping:

\begin{equation}
 \tau_\mathrm{Pr}=2\sqrt{5\pi}\frac{2R}{\delta v} \frac{\rho_i+\rho_e}{\rho_i-\rho_e}=2\sqrt{5\pi} \frac{P}{2\pi a} \frac{\rho_i+\rho_e}{\rho_i-\rho_e}, 
 \label{eq:uni_turb} 
\end{equation}
with $\delta v$ the velocity amplitude of the kink mode. The presence of  $\delta v$ in Eq. \ref{eq:uni_turb} highlights the dependence on the kink mode amplitude. The second form of the time-scale is written in terms of a normalised amplitude, $a=A/R$, given that $\delta v=2\pi A/P$ (where $A$ is the displacement amplitude and $P$ the wave period). The exact interplay  between the resonant absorption and uni-turbulence mechanisms in kink wave damping is still unclear.
There is a likely competition between both methods in damping the kink modes, with uni-turbulence playing a dominant role when the wave amplitude is significantly large. As of yet, a full comparative study does not exist.

\subsection{Observed wave damping}

While it is currently somewhat difficult to measure the development of small-scales and the subsequent dissipation of wave energy, an estimate of the rate of damping of the kink waves is currently feasible.

There are now many observational examples of the damped standing kink mode in active region loops \citep[e.g.,][]{ASCetal2002, veretal2004}, which appear well described by resonant damping \citep[e.g.,][]{GOOetal2002,RUDROB2002,Aschwanden_2003}.
\citet{GODetal2016} has complied a catalogue of damped standing kink waves, which was later extended by \citet{Nechaeva_2019} to include 223 oscillation events. From this catalogue, the rate of wave damping ($\tau_\mathrm{RA}/P$) is typically $< 10$, with a mean of 2.3 (see 
Section~\ref{sec:obs} for details). \cite{GOONAK2016} presented a meta-analysis of a number of standing kink modes cases, which suggested there was a dependence on the 
damping rate with respect to the amplitude of the oscillation. Given resonant damping is independent of amplitude (see Eq.~\ref{eq:equib}), the results could indicate the action of a non-linear damping mechanism for large amplitude kink waves. 

Relevant to our work, \citet{VERetal2013b} demonstrated the potential of utilising statistical approaches that forward model the damping of kink waves in order to constrain the cross-sectional characteristics of the coronal loops. The essence of their approach was to generate physical properties of the coronal loop (e.g. density contrast, length-scale of inhomogeneous layer) by randomly drawing values from assumed distributions, which were fed through analytical expressions for resonant damping. This enabled synthetic distributions of damping times to be generated that could be compared to an empirical distribution of 50 existing cases. Distributions of loop parameters were optimised to improve the fit between synthetic and empirical distributions.  

Recently, \citet{Van_Doorsselaere_2021} took a similar approach, although focused on the modelling of the non-linear damping of the standing kink waves (incorporating an analytic expression for the non-linear 
damping, i.e., Eq.~\ref{eq:stand}). The damping times were matched with the simulation results of 
\citet{magyar2016} and the empirical damping times from the catalogue 
of \citet{Nechaeva_2019}. They obtaining a good agreement between the Monte Carlo simulations of damping times in loop models and the observations, suggesting that the observed amplitude-dependent damping to be a consequence of the $A^{-1}$ scaling present in the expression for damping rates (see Eqs.~\ref{eq:uni_turb} \& \ref{eq:stand}).

\medskip

Up until now, the observations of damped propagating coronal kink waves were fewer. In fact, they have been restricted to a single case
observed with CoMP that was originally reported by \citet{TOMMCI2009}. \citet{VERTHetal2010} demonstrated the damping for the event was frequency-dependent, and noted the observed behaviour was described 
well by resonant absorption \citep{TERetal2010c}. Further, \citet{VERTHetal2010} suggested the ratio of the power for outward to inward propagating kink waves, namely
\begin{equation}
    \langle P(f)\rangle_{ratio} = \frac{P_{out}}{P_{in}}\exp\left(\frac{2L}{v_{ph}\xi_\mathrm{E}}f \right),
    \label{eq:pow_rat}
\end{equation}
should provide a reasonable model in order to measure the damping rate for propagating waves. Here $L$ is the loop length, $\xi_\mathrm{E}$ is referred to as the quality factor or equilibrium parameter, $v_{ph}$ is the phase speed, $P_{out}$ is the power of waves propagating upwards and $P_{in}$ is the power of the waves propagating inwards. This expression\footnote{Equation~\ref{eq:pow_rat} also assumes that the power ratio is averaged along half of the coronal loop, \cite{TIWARI_2019} provide an alternative for this expression when only a section of the loop is used to measure the power ratio.}  is general and applicable for any mechanism where the damping is exponential in nature \citep[c.f.][for discussion of the Gaussian damping regieme in resonant absorption]{HOOetal2013}. Other studies have also used this single example of the damped propagating kink wave, interpreting the observed frequency-dependent damping as resonant absorption \citep{VERetal2013b,PASetal2015,Montes_Solis_2020}. Recently, \citet{TIWARI_2019} demonstrated that some of these previous studies likely underestimate the damping length by erroneously assuming the generative distribution for the power ratio is described by a Gaussian distribution (applicable when fitting Eq.~\ref{eq:pow_rat} to the power spectrum obtained from CoMP data). 

\medskip

The preceding discussion highlights the need to measure the rate of damping for propagating kink waves. It is a key piece of 
information required to substantiate the role of resonant absorption, phase mixing and/or uni-turbulence in the heating of the quiescent corona. Such measurements 
place a limit on the rate of energy transfer from the global kink modes to the smaller scales, subsequently influencing the rate of heating. While only suggestive, \citet{TIWARI_2019} found that the quality factor for quiescent coronal loops was large compared to the typical values found in active region loops. Furthermore, the damping rates can also be used for magnetoseismology, enabling the cross-sectional properties of coronal loops to be inferred. Here we extend this previous analysis and present the key result of a larger study of propagating kink waves in quiescent coronal loops.

\section{Observations and data analysis}\label{sec:obs}

All data used within this study were taken with the CoMP instrument between the dates of 20 January 2012 and 02 January 2014. Details of individual data sets used are given in a companion publication, \citet{Tiwari_2021}, which discusses the catalogue of events in more detail. The data were processed using the standard CoMP reduction pipeline and the Doppler velocity products are utilised. Additional 
registration of the data with cross-correlation is undertaken to achieve sub-pixel alignment of time-sequences for individual days. 
Full details of the registration process is discussed in \cite{MORetal2016}. The subsequent analysis of the data is described fully in \citet{TIWARI_2019, Tiwari_2021}, although we give a brief overview 
here.

We restrict our attention to quiescent coronal loops, avoiding active regions. In order to minimise geometric influences on the measurements, e.g. foreshortening, the coronal loops that are selected for analysis are orientated such that the longitudinal axis is close to being in the plane-of-sky.  Time-distance diagrams are obtained that follow the direction of wave propagation along the coronal loops \citep[e.g.,][]{TOMMCI2009, MORetal2015, TIWARI_2019}. 
For each of the time-distance diagrams, a two-dimensional Fourier transform is applied to separate the outward and inward propagating wave components. The power spectra in frequency--wavenumber space is then obtained. Summing over wavenumber and taking the ratio of the outward to inward waves provides the power ratio as a function of frequency. The power ratio is then fit with the exponential model given in Eq.~\ref{eq:pow_rat} using maximum likelihood, assuming the power ratio is distributed about the true value following an \emph{F}-distribution \citep[see][for full details of the maximum likelihood approach]{TIWARI_2019}.

\begin{figure}[!t]
\includegraphics[scale=0.55, clip=true, viewport= 15 0 500 320]{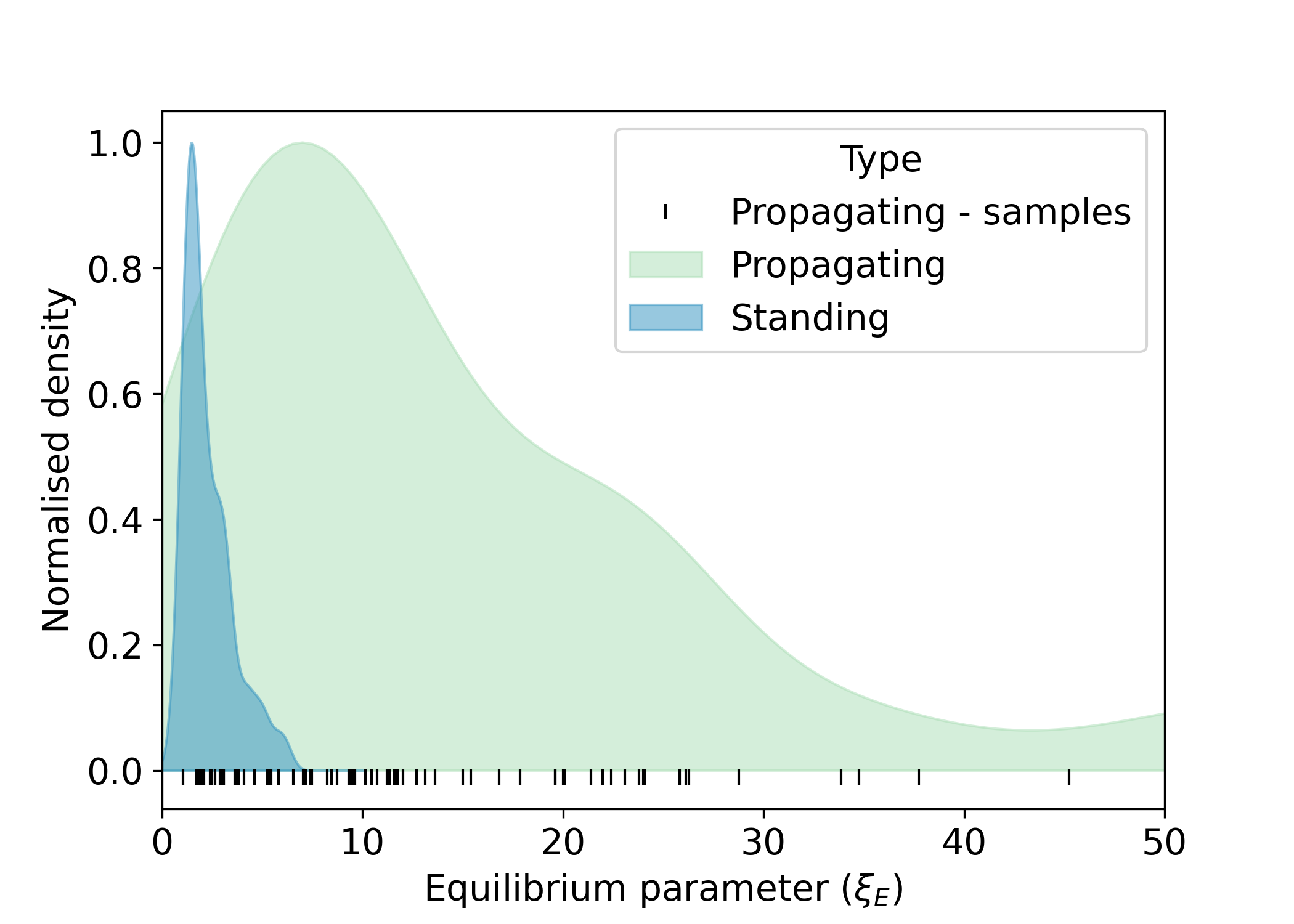}
\caption{Distribution of equilibrium parameter for \emph{propagating} kink waves in 77 quiescent coronal loops. The plot shows
the kernel density estimation for the distribution, with the individual samples indicated by the vertical marks at the bottom of the plot. Six values for $\xi_\mathrm{E}$ are greater than 50 and are not shown in the graph for visualisation purposes. For comparison, the distribution of $\xi_\mathrm{E}$ for damped \emph{standing} kink waves measured in active regions is also shown. The two density distributions are both normalised to further aid visualisation.}\label{fig:equil}
\end{figure}

In order to estimate the equilibrium parameter, $\xi_\mathrm{E}$, for each loop, we are required to measure the propagation speed of the waves. This is achieved via a coherence analysis of the velocity time-series that make up each time-distance diagram \citep{TOMMCI2009,TIWARI_2019}. In all, we provide estimates for the rate of damping of propagating kink waves in 77 quiescent coronal loops, which is shown in Figure~\ref{fig:equil}. The density distribution of the equilibrium parameter is estimated with Kernel Density Estimation (KDE) using a Gaussian kernel. The band-width parameter is selected via cross-validation\footnote{Performed with Scikit-learn \citep{scikit_learn}.}.

To provide some perspective for the equilibrium parameters from propagating kink waves in quiescent coronal loops, we also calculate the equilibrium parameter for the damped standing kink waves observed in coronal loops in active regions. The data for the active region coronal loops and wave parameters are taken from the catalogue of standing kink modes given in \citet{Nechaeva_2019}. The equilibrium parameter for the standing modes is calculated by taking the ratio of the estimated damping time by the period. The results are shown in Figure~\ref{fig:equil}. Catalogue entries without a period or damping time estimate are excluded, leaving 101 samples.

\section{Weak damping in the quiescent Sun}

The most significant conclusion that can be drawn from the distribution of equilibrium parameter (Figure~\ref{fig:equil}) is that the propagating kink waves
in the quiescent Sun are weakly damped. This conclusion is made starkly clear in comparison to the
corresponding values for the (rapidly-damped) standing kink modes. The values of the equilibrium parameter for the propagating waves are in the range $0.9<\xi_\mathrm{E}<298$. This range is significantly broader than that of the equilibrium parameters
for standing waves ($0.7<\xi_\mathrm{E}<7$). The central tendencies of the two sets of samples are also different, with the propagating waves having a mean value of $24\pm4$ (median 11) and the standing waves have a mean $2.3\pm0.1$ 
(median 1.8); the uncertainties given are standard errors.

Assuming that the dominant mechanism for the observed kink wave damping is resonant absorption, the implication of the large equilibrium parameter values is that 
the quiescent loops must be physically different to those found in active regions. This is perhaps not such a surprise. From Eq.~\ref{eq:equib}, it is seen that 
the value of the equilibrium parameter is determined by the difference in densities between the internal and external plasma (although not the magnitude of the 
density). Consequently, the density ratio of loop plasma to ambient plasma ($\rho_i/\rho_e$) is 
smaller in quiescent loops, relative to active region loops. 

The other quantity in Eq.~\ref{eq:equib} is the relative size of the inhomogeneous boundary layer, $l/R$, which results from an approximation to the gradient of
density across the inhomogeneous layer, namely $\mid d\rho/dr \mid$ \citep[e.g.][]{GOO2008}. It was shown by \cite{ARRGOO2019} that the variation of density across the inhomogeneous layer can be factored into a single quantity (which they denote $G$) and is essentially a shape factor that \lq\lq\textit{relates the local variation of density at the resonant position to the global variation of the density over the mean radius of the cylinder}\rq\rq. The overall value of this quantity $G$ can influence the size of $\xi_\mathrm{E}$, and it likely takes a range of values across different coronal structures. However, we see no reason why the distribution of $G$ values would be different for quiescent and active region loop populations. 

In our opinion, this places a focus on the density ratio as the underlying factor determining the 
differences in the strength of damping between the quiescent loops and active region loops. Hence, the inference is that active region loops are denser than quiescent loops, relative to the ambient plasma. This has support from work on coronal loop thermodynamics. Under the uncontroversial assumption that active region loops have a greater heating rate than quiescent loops, active region loops are expected to be denser \citep[see, e.g. the scaling law Eq.~45 for coronal densities in ][]{Bradshaw_2020}.  

A similar inference can be drawn if uni-turbulence dominates the damping. Eq.~\ref{eq:uni_turb} shows that the rate of energy transfer has the same 
dependence on the density contrast. However, the damping rate is also influenced by $R/\delta v$. 

In order to examine the role of the density contrast and 
the roles of resonant absorption/uni-turbulence further, we undertake a Monte Carlo forward-modelling approach. The general approach follows that of \cite{VERetal2013b} and \citet{Van_Doorsselaere_2021}, where  values for
$\xi_\mathrm{E}$ are generated by the random sampling of coronal loop parameters and observed wave parameters, then feeding those values through analytical expressions of damping rates. The following subsections provide the details for the modelling.

\subsection{Assumptions and equations}
The models that describe the physics within the coronal loops are the same as that in \citet{Van_Doorsselaere_2020} and \citet{Van_Doorsselaere_2021}. The coronal loops are 
assumed to consist of a straight cylinder with radius $R$ embedded in a uniform magnetic field with a low-$\beta$. The internal density $\rho_\mathrm{i}$ is higher than 
the external density $\rho_\mathrm{e}$, resulting in a density contrast $\zeta=\rho_\mathrm{i}/\rho_\mathrm{e}$. 
The loop is oscillating with a period $P$ and a displacement amplitude $A$.

The damping for the propagating waves by uni-turbulence is given in Eq.~\ref{eq:uni_turb}, while the damping for standing waves by the 
KHI development, $\tau_\mathrm{St}$, was derived in \citet{Van_Doorsselaere_2021} (taking into account the
erratum) and is given by:
\begin{equation}
    \tau_\mathrm{St}=40\sqrt{\pi}\frac{P}{2\pi a}\frac{1+\zeta}{\sqrt{\zeta^2-2\zeta+97}}.
    \label{eq:stand}
\end{equation}

Furthermore, we utilise the expression for the damping time of resonant absorption, $\tau_\mathrm{RA}$, for standing modes given by \citet{RUDROB2002}, which was derived in 
the thin-tube, thin-boundary limit, and which assumed a sinusoidal density profile in the inhomogeneous layer\footnote{ 
It is worth mentioning that the formulae for the non-linear damping $\tau_\mathrm{St,Pr}$ considers a step function in density instead.}  with width $l$. This expression is similar to Eq.~\ref{eq:equib}, with the left-hand-side equal to $\tau_\mathrm{RA}/P$.  For propagating waves, we consider Eq.~\ref{eq:equib} and assume that damping length and damping time are closely related:
\begin{equation}
    \xi_\mathrm{RA}=\frac{L_\mathrm{D}}{\lambda}=\frac{\tau_\mathrm{RA}}{P},
\end{equation}
which is correct for weak damping, but not true in general in complicated dispersion relations. Here we denote the equilibrium parameter for resonant absorption as $\xi_\mathrm{RA}$. 

Following the approach of \citet{Van_Doorsselaere_2021}, we combine the damping of resonant absorption with the non-linear damping through the formula:
\begin{equation}
    \exp{\left(-\frac{t}{\tau}\right)}=\exp{\left(-\frac{t}{\tau_\mathrm{St,Pr}}\right)}\exp{\left(-\frac{t}{\tau_\mathrm{RA}}\right)},
\end{equation}
allowing us to express the total damping with $\tau$ as the harmonic average:
\begin{equation}
    \frac{1}{\tau}=\frac{1}{\tau_\mathrm{St,Pr}}+\frac{1}{\tau_\mathrm{RA}}.
    \label{eq:combine}
\end{equation}
Taking the harmonic average is a typical approach if the damping of two mechanisms is small and independent (and thus additive). This is likely satisfied for the propagating waves, but the assumptions are violated for the standing modes where there is strong damping (and the presence of resonant absorption kick-starts the non-linear 
damping, see \citealt{antolin2019}). Still, it is a first-order approach to the combination of both damping mechanisms, and needs revision in the future.

\par
{
In the following, we use the equilibrium parameter, rather than the damping time, as it is independent of frequency for both resonant absorption and the non-linear damping. Multiplying Eq.~\ref{eq:combine} by the period, we define the total equilibrium parameter as
\begin{equation}
    \frac{1}{\xi_\mathrm{E,Total}}=\frac{1}{\xi_\mathrm{NL}}+\frac{1}{\xi_\mathrm{RA}}.
    \label{eq:combine_2}
\end{equation}
The equilibrium parameter for the non-linear damping is denoted $\xi_\mathrm{NL}$.
}

\begin{table}[!t]
\caption {Physical Parameters for forward-modelling}
\begin{tabular}{ l c  c }

  \hline
  Parameter & Range & Distribution\\
  \hline            
  $\zeta$ standing & 1-5 & $\mathcal{U}(1,5)$ \\
  $\zeta$ propagating & 1-1.3 & $\mathcal{U}(1,1.3)$ \\
  $l/R$ & 0-2 & $\mathcal{U}(0,2)$ \\
  Loop Radius ($R$) & 0.5-10~Mm & $\mathcal{U}(0.5,10)$ \\
  \hline  \label{tab:phys_param}
\end{tabular}
\end{table}

\subsection{Physical parameters}
In order to simulate the damping rates using the above equations, we need to specify the latent values for the density contrast, width of the inhomogeneous layer 
and loop radius\footnote{In principle one could measure the loop radii for the standing modes with AIA, but we do not undertake that here. In the CoMP data, the velocity signal is averaged over neighbouring loops due to spatial resolution; hence a single radius can not be associated with the observed kink waves.}.  Moreover, we also require the typical amplitudes of oscillations. We discuss our choices next and a summary is given in Table~\ref{tab:phys_param}.

\medskip
\noindent\textit{Propagating wave amplitude ($A$)} While we are able to estimate the kink wave damping with CoMP, the limited spatial resolution of the instrument
means that observed wave amplitudes are reduced due to spatial averaging and line-of-sight integration \citep[see e.g.,][]{DEMPAS2012,MCIDEP2012,Pant_2019}. Hence,
we use measurements of the transverse displacements of loops in the quiescent Sun observed with the SDO Atmospheric Imaging Assembly \citep[AIA -][]{LEMetal2012}  
in the $171$~{\AA} channel, 
taking the values of amplitude discussed in \cite{MORetal2019}. The data set contains 590 values.

\medskip
\noindent\textit{Standing wave parameters ($A$)} The standing kink mode amplitudes are taken from the catalogue of \citet{Nechaeva_2019}, which represents 
kink waves in active region coronal loops measured with SDO/AIA. The catalogue contains 223 values.

\medskip

\begin{figure*}[!th]
\centering
    \includegraphics[scale=0.48, clip=true, viewport= 15 0 500 350]{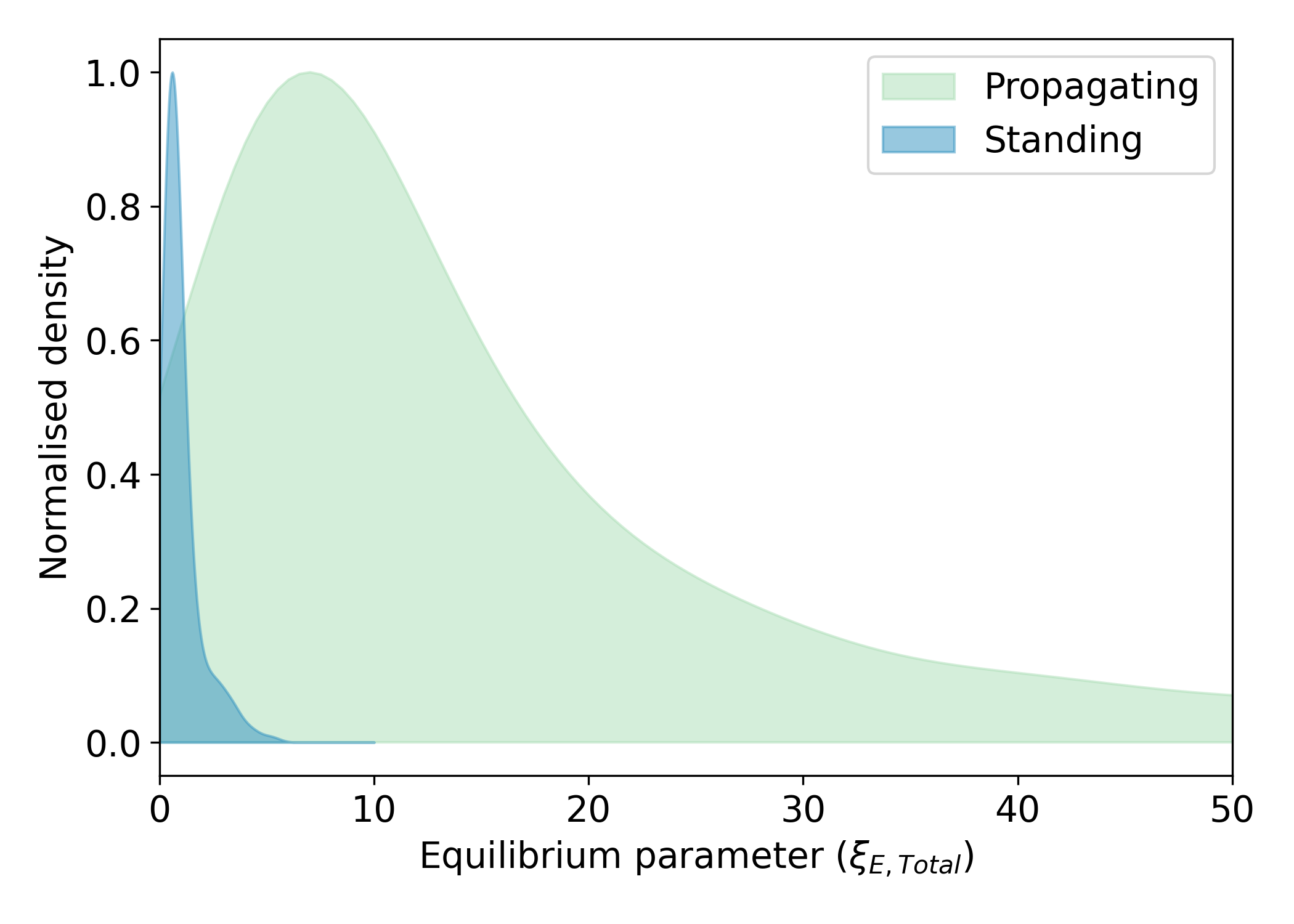}
    \includegraphics[scale=0.48, clip=true, viewport= 15 0 500 350]{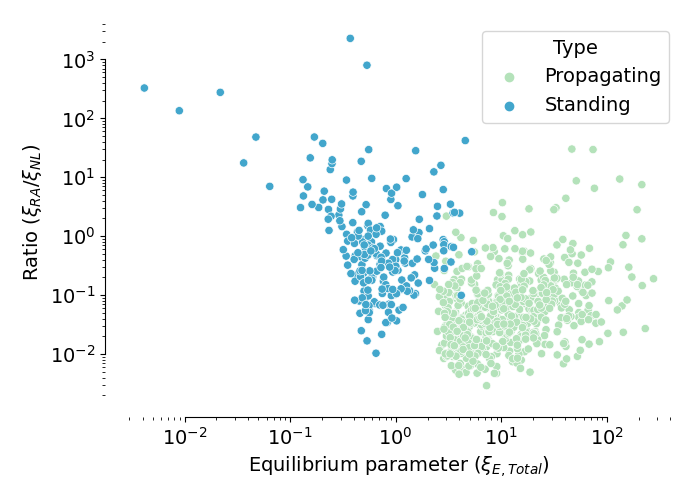}
    \caption{Results from the Monte Carlo simulations. The left panel shows the simulated distributions of equilibrium parameters. The right panel reveals the contributions of resonant absorption ($\xi_\mathrm{RA}$) and non-linear damping ($\xi_\mathrm{NL}$) in the overall damping ($\xi_\mathrm{E,Total}$) of both wave modes.  }
    \label{fig:simulations}
\end{figure*}

\noindent\textit{Loop Radii ($R$)} The radius of coronal loops is a well discussed topic. Past investigations have focused largely on active region coronal loops, in part due to their high visibility in coronal EUV images. \citet{ASCPET2017} provide a detailed investigation of active region 
coronal loop widths using AIA and Hi-C, along with a summary of all previous observations to date \cite[results from the more recent Hi-C 2.1 mission can be found in][and are in agreement]{Williams_2020}. They report that the widths of coronal loops peak at 0.5~Mm but can be as large as 10~Mm. However, the resolved loops observed with AIA have a larger
peak at $\sim1.2$~Mm, which may better represent the coronal loops in which the standing kink modes are found. To date, there are no similar measurements for quiescent loops 
and for now we assume that the widths are similar. For this study, we assume the loop widths are uniformly distributed with $R=[0.5,10]$~Mm. The impact of choosing a uniform
distribution over this range means that we are selecting larger loop radii more often than observed. This will lead to an underestimate of the strength of non-linear damping.

\medskip

The remaining parameters are latent variables in the model, namely $\zeta$ and $l/R$. There have been numerous efforts in the past to determine the size of the density contrast and 
the width of the inhomogeneous layers in coronal loops using coronal seismology \citep[e.g.][]{Aschwanden_2003,Arregui_2011,Asensio_Ramos_2013,VERetal2013b,Goddard_2017,Pascoe_2018}. The focus of these attempts has exploited only standing modes in active region loops, and so the estimated values of the density contrast naturally 
reflect the conditions of active region loops. Moreover, each of them has assumed only resonant absorption is in action, which will likely influence the 
estimated values.

\medskip
\noindent\textit{Width of inhomogeneous layer ($l/R$)} The inhomogeneous layer can be in the range $l/R=[0,2]$ and the above studies generally favour larger values
of the inhomogeneous layer, suggesting the resonant layer is thick. We take a more conservative approach and assume the values are drawn uniformly from the range.
If the true values of $l/R$ are typically towards the larger end of this range, the impact of our choice will be to underestimate the strength of the resonant damping.

\medskip
\noindent\textit{Density contrast ($\zeta$)} 
 The results across the previous studies are relatively consistent, with the majority of values in the range $\zeta=[1,10]$. However, the estimates tend
to have a preference for small values ($\zeta<5$). Hence, in the active region loop model we choose to draw values uniformly in the range $\zeta=[1,5]$. 

For the coronal loops in the quiescent Sun, which support the propagating kink waves, to our knowledge there have been no previous estimates. It might be natural to expect that the 
density of the quiescent loops is closer to the ambient coronal value for two reasons. The first is that the quiescent loops visible contrast in EUV images is lower than their active region counterparts, suggesting similar densities and temperatures to the ambient corona. Secondly, the required heating rate is at least five times smaller in the quiescent Sun. This would imply that the energy released (through whichever mechanism) is less and, as such, there is less energy available to contribute to ablating the 
Transition Region and chromosphere
and reducing evaporation of the denser plasma into the loops \citep[e.g.][]{Cargill_2012}. Given the unknown nature of $\zeta$ in the quiescent Sun, we leave this as 
a free parameter that we vary in order to obtain a reasonable match between simulations and the observed data.

\subsection{Monte Carlo approach}

{For each of the observed cases (both propagating and standing) we select a value of wave amplitude. This is de-projected by using a randomly drawn angle between $\phi=[-\pi/2+\delta\phi,\pi/2-\delta\phi]$ to compute 
$A_{de-project}=A/\cos{\phi}$, with the assumption that the oscillation plane is oriented randomly with respect to the line-of-sight. The $\delta\phi$ is intended to represent a limit on the orientation of the displacement, beyond which the oscillation is essentially along the line-of-sight and cannot be measured by imaging instruments. Here we set $\delta\phi=\pi/18$.
We also take a realisation of width of inhomogeneous layer ($l/R$), density contrast ($\zeta$) and radius ($R$) 
drawn from the appropriate distributions.}  

The only free parameter that we have in models is the range of $\zeta$ for the quiescent coronal loops. As mentioned there are no previous measurements to guide us. Hence,
we vary the upper value of $\zeta$ until various moments and statistics of the observed and simulated distributions are in rough agreement. In particular we examined the mean, median, mode, variance, 
skew and kurtosis. This led to the range $\zeta=[1,1.3]$ and the resulting distributions for $\xi_\mathrm{E, Total}$ are shown in Fig.~\ref{fig:simulations}. {The values of the moments/statistics are given in Table~\ref{tab:dist_vals} for the observed distribution and also for the Monte-Carlo results for the given range of $\zeta$ (means and standard deviations of values calculated from a 1000 different simulations). For reference, increasing the maximum value of $\zeta$ to 2 leads to significantly smaller values of mean, median and mode than observed, and larger values of skew and kurtosis. We note that the range of $\zeta$ is not optimized but chosen by hand, and could probably be more finely tuned by minimising some objective function that evaluates a measure of distance between the observed and simulated distribution. However, our goal here is to demonstrate the idea that, assuming resonant and/or non-linear damping, density contrast is responsible for the larger values of equilibrium parameter, rather than tune the model parameters.}

\begin{table}[!t]
\caption {Observed vs simulated distributions of $\xi$ for propagating waves.}
\begin{tabular}{ l c  c }

  \hline
   & Observed & Monte-Carlo \\
  \hline            
  Mean & 24 & $23\pm1.5$ \\
  Median & 11 & $10\pm0.6$\\
  Mode & 5 & $7.0\pm0.3$\\
  Std. Dev. & 39 & $35\pm3$ \\
  Skew & 4.6 & $3.7\pm0.3$ \\
  Kurtosis & 27 & $17\pm3$\\
  Max value & 298 & $275\pm22$\\
  Min Value & 1.05 & $2.0\pm0.3$\\
  \hline  \label{tab:dist_vals}
\end{tabular}
\end{table}

\subsection{Results}
Figure~\ref{fig:simulations} shows the results from the forward-modelling. We first focus on the distribution of the simulated equilibrium parameter, ($\xi_\mathrm{E, Total}$ - shown in the left panel), which we assume is comparable to the observed values of $\xi_\mathrm{E}$. Comparing Figures~\ref{fig:simulations} and \ref{fig:equil} reveals a remarkable agreement for propagating waves. Even with relatively conservative assumptions about key physical parameters and the ability to `tune' one parameter, the model can provide a reasonable description of the observed distribution of the  equilibrium parameter. We are able to match the aforementioned moments and statistics of the observed distribution reasonably well (Table~\ref{tab:dist_vals}), hence the
visible agreement between density distributions. {We stress that we are not suggesting that, in the quiescent corona, the largest value of $\zeta=1.3$. The modelling used a uniform distribution for $\zeta$ but it is possible that a skewed, long-tailed distribution for $\zeta$ (i.e., with occasional larger values of $\zeta$) could also be feasible. }

The simulated distribution for the standing modes is narrower with smaller measures of central tendency, i.e., the model provides equilibrium parameters that are smaller than observed. This is likely due to the assumptions in the model not holding in the strong damping regime. However, we do not focus on these differences further given our focus is the propagating waves. 

\medskip
It is also informative to look at the role the two different damping mechanisms play in the total damping. In order to investigate this, we show the ratio of the resonant damping equilibrium parameter 
to the non-linear damping equilibrium parameter, i.e., $\xi_\mathrm{RA}/\xi_\mathrm{NL}$. This quantity is compared to the total equilibrium parameter in the right hand panel of Figure~\ref{fig:simulations}. It can be seen for the propagating waves that, generally, $\xi_\mathrm{RA}/\xi_\mathrm{NL}<1$, which means that resonant damping dominates over non-linear damping. This result will depend on the wave amplitude, which in turn depends on the frequency \citep{MORetal2019}. The situation is different for standing modes, where non-linear damping appears to play an equivalent or greater role in the overall wave damping.

\section{Discussion}

The damping of propagating kink waves in the quiescent corona is found to be weaker than the damping of the standing kink modes in active regions. Both our intuition and the results of the forward modelling support the role of the density contrast playing a key role in the damping of 
kink waves in the quiescent Sun. We note that the forward-modelling assumes the other physical aspects of coronal loops, i.e., transverse homogeneity length scale and radius, are similar for active region and quiescent loops. Of these three parameters, we can only motivate a physical reason for the density contrast differing between the two sets of loops.

We believe it is also worth mentioning the observations of propagating kink waves along magnetised plasma structures in coronal holes. While the structures in coronal holes are not loops, they still appear (marginally) brighter in EUV emission than the ambient plasma \citep[e.g.,][]{MORetal2015}; implying they are also over-dense, although less-so than active region loops and quiescent loops. This property is reflected in the observed wave behaviour. It has been shown in \citet{MORetal2015} and \citet{Weberg_2020} that the propagating kink waves in coronal holes demonstrate an increase in amplitude that is consistent with expectations from WKB wave propagation in an gravitationally-stratified plasma 
(at least below 1.3~$R_{\odot}$), i.e., there is no evidence for wave damping\footnote{This wave amplification is also found in measurements of non-thermal line widths below 1.3~$R_{\odot}$ \citep[e.g.,][]{BANetal1998,BEMABB2012,HAHetal2012}; with the large non-thermal line widths thought to be due to the kink waves \citep{MCIDEP2012,DEMPAS2012,Pant_2019}. The apparent wave damping above 1.3$R_\odot$ reported by \citet{BEMABB2012} and \citet{HAHetal2012} is still somewhat a mystery, and could be evidence of non-WKB propagation (\citealt{PANT_2020}) rather than actual wave dissipation.}. 

Hence, if the density contrast between the wave-guide and the ambient plasma is the main factor behind the differences in observed damping rates between standing kink modes in active regions and propagating kink waves in the quiescent corona 
(and in coronal holes), this would indicate the magnetised wave-guides display a spectrum of density ratios. The largest density ratios occur in active region loops (leading to the rapid observed damping of the kink waves), and the smallest in coronal hole wave-guides (leading to no observable wave damping).

\medskip

The arguably more significant implication of the observed weak damping is that the rate of energy transfer from the global kink motion to azimuthal motions is low. This potentially rules out the energy pathway of resonant absorption to phase mixing as a viable mechanism for dissipating the kink wave energy in the quiescent corona. The numerical studies of \cite{Pagano_2017,Pagano_2019,Pagano_2020} appear to show that phase mixing is inefficient for heating even when the rate of energy transfer is likely at its largest, i.e., under active region conditions. Moreover, the rate at which small-scales develop due to phase mixing is governed by the density gradient through the inhomogeneous boundary layer. This can be seen in an approximate expression for the length scale due to phase mixing, $L_{pm}$, is given by \cite{Mann_1995}:
\begin{equation}
L_{pm} = \frac{2\pi}{\mid d\omega_a(r)/dr\mid t},    
\end{equation}
where $\omega_a(r)$ is the Alfv\'en frequency across the inhomogeneous boundary layer\footnote{This formula has been shown to give a reasonable description of the development of phase mixing length scales in both analytical and numerical treatments of the resonant absorption of kink modes \citep{PASetal2013,SOLetal2015}.}. Hence, smaller density ratios in the quiescent coronal loops (relative to active region loops), will not only reduce the rate of transfer of energy from the global kink mode to the Alfv\'en modes, but also delay the development of the small-scales due to phase mixing.

The role of the non-linear damping due to uni-turbulence appears to be less than resonant damping (as suggested from the forward-modelling), so this will also play a small role in wave damping in the quiescent Sun.

\medskip

The observed weak damping and implication of inefficient energy transfer to small spatial scales means that we currently find it difficult to see how phase mixing can play an important role in dissipating kink wave energy in the quiescent corona. This result would naturally extend to the wave-guides in coronal holes also. However, this does not rule out the propagating kink waves as a key player in the heating of the quiescent corona. There are
still a number of potential mechanisms that might be able to dissipate the wave energy. A leading candidate from wave-based heating is Alfv\'enic turbulence, which has had some success in producing heated coronal loops \citep{Buchlin_2007, VANBALLetal2011,van_Ballegooijen_2017} and also in producing a hot, fast solar wind \citep[e.g.,][]{CRAetal2007,Shoda_2018,Shoda_2019}. The current models ignore the perpendicular density structure so do not contain kink modes, however the Alfv\'enic nature of the low-frequency kink modes suggest they could also be susceptible to the same non-linear interactions that occur between counter-propagating waves and transfer energy to smaller scales. We eagerly await the results of an Alfv\'enic turbulence model that would include the structured nature of the lower solar corona and would shed light on the dissipation of the wave energy in kink modes.

\medskip

{Finally, the weak damping rates estimated from this work raise a puzzling question; \textit{why do we not see any evidence for resonant frequencies in the loops studied in the CoMP data?}  The standard picture of wave propagation through the Sun's atmosphere suggests 
that the Alfv\'enic waves in the corona will be reflected off the Transition Region due to steep gradients in Alfv\'en speed 
that is assumed to exist there (due to a significant change in density over a relatively short length-scale). Resonances are 
expected to occur at frequencies of $f=nv_A/2L$, where $n=1,2,...$ \cite[e.g.,][]{ION1982,Holl1984}. The typical length of the loops in which we analyse the wave events are $\sim400$~Mm  with Alfv\'en speeds of $\sim500$~km/s \citep{Tiwari_2021}. This 
leads to Alfv\'en travel times $\tau_A\approx 800$~s, whereas the time-scale related to the wave damping is $\tau_D\approx 3300$~s (for a wave with period 300~s). Hence, in principle the waves should be able to traverse the loop twice before any significant damping occurs, and set up resonant frequencies at $f=0.625n$~mHz (a number of which should be observable by CoMP if they were present). In contrast, the decay-less kink oscillations
found in active region coronal loops are thought to be the resonant 
oscillations driven through random footpoint motions \citep{Hindman_2014, Afanasyev_2020} and shown to be standing modes by \cite{ANFetal2013}. } 

{There are at least two factors at work that may help explain the lack of observed loop resonances, and are likely acting in concert. The first is that there will likely be some turbulent dissipation of the waves \cite[e.g.,][]{MATetal1999,Dmitruk_2002,VANBALLetal2011,Verdini_2012}, given that there exists counter-propagating Alfv\'enic waves along each of the loops. If we are observing the frequencies/wavelengths associated with the inertial range, then energy transfer would be self-similar and hence is independent of frequency, i.e., with no impact upon the measurement of frequency dependent damping. Secondly, there should be some degree of transmission (or leakage) of the waves in the corona to the chromosphere. In principle, both of these phenomenon should not, however, inhibit the development of resonances, but would play a role in limiting the amount of energy that can build up in the coronal volume \citep{Verdini_2012}.} 

{As noted, previous work that has discussed the development of resonances in coronal loops generally assumes a steep gradient in the Alfv\'en speed at the Transition Region or splits the atmosphere into cavities with jumps in Alfv\'en speed \cite[e.g.][]{ION1982,Holl1984,Verdini_2012}. However, it is not evident that this is the case in the quiescent Sun. The notion of a steep density gradient (and hence Alfv\'en speed gradient) across the Transition Region is built upon the premise that atmosphere is in hydrostatic equilibrium. Recent numerical and observational work suggests this is far from the case. 
The lower atmosphere is known to be replete with shocks \citep[e.g.,][]{LEEetal2007,LEEETAL2012} and spicules 
\citep[e.g.,][]{PERetal2012}, which are continually driving chromospheric material into the corona and potentially removing the steep 
density gradient \citep[e.g.,][]{MARetal2017}. The time-averaged profile of the density across the Transition Region could well be different from the classical hydrostatic case, with tentative evidence for this in coronal holes \citep{Weberg_2020}. The non-hydrostatic nature of the lower atmosphere could well lead to significantly less wave reflection and prevent the 
existence of resonant frequencies in the quiescent corona. In 
contrast, active regions are known to contain much shorter jets, known as dynamic fibrils \citep{DEPetal2007c}, which would not spread 
the chromospheric material as thinly and potentially leave the Alfv\'en speed jump at the Transition Region intact.} 
  
{It is clear there are still a number of questions remaining about the structure of the solar atmosphere and its impact on wave propagation. We believe incorporating the dynamic nature of 
chromosphere and Transition Region is an important feature for examining wave propagation and subsequent heating in the quiet Sun.}

\begin{acknowledgments}
We are grateful to the anonymous referee whose comments helped improve the manuscript. R.J.M. is supported by a UKRI Future Leader Fellowship (RiPSAW - MR/T019891/1), and thanks R. Soler and I. Arregui for providing comments on a draft of the manuscript. A.K.T is supported by the European Union's Horizon 2020
research and innovation programme under grant agreement No 824064 (\href{https://projectescape.eu/}{ESCAPE}). T.V.D. was supported by the European Research Council (ERC) under the European Union's Horizon 2020 research and innovation programme (grant agreement No 724326) and the C1 grant TRACEspace of Internal Funds KU Leuven. J.A.M. is supported by the Science and Technology Facilities Council (STFC) via grant number ST/T000384/1. The authors also acknowledge STFC via grant number ST/L006243/1 and for IDL support. The CoMP data is courtesy of the Mauna Loa Solar Observatory, operated by the High Altitude Observatory, as part of the National Center for Atmospheric Research (NCAR). NCAR is supported by the National Science Foundation. 
\end{acknowledgments}
%

\vspace{5mm}
\facilities{Solar Dynamics Observatory/Atmospheric Imaging Assembly, CoMP}

\software{ NumPy \citep{Numpy}, matplotlib \citep{Matplotlib}, seaborn \citep{seaborn}, Scikit-learn \citep{scikit_learn}, SciPy \citep{Scipy}, IPython \citep{IPython}.}


\pagebreak



\bibliographystyle{aasjournal}

\end{document}